# A Stability Formula for Plastic-Tipped Bullets

## Part 2: Experimental Testing

Michael W. Courtney, Ph.D., U.S. Air Force Academy, 2354 Fairchild Drive, USAF Academy, CO, 80840
Michael_Courtney@alum.mit.edu

Donald G. Miller, Ph.D., 2862 Waverley Way, Livermore, CA 94551

**Abstract:** Part 1 of this paper describes a modification of the original Miller twist rule for computing gyroscopic bullet stability that is better suited to plastic-tipped bullets. The original Miller twist rule assumes a bullet of constant density, but it also works well for conventional copper (or gilding metal) jacketed lead bullets because the density of copper and lead are sufficiently close. However, the original Miller twist rule significantly underestimates the gyroscopic stability of plastic-tipped bullets, because the density of plastic is much lower than the density of copper and lead. Here, a new amended formula is developed for the gyroscopic stability of plastic-tipped bullets by substituting the length of just the metal portion for the total length in the *(1 + l²)* term of the original Miller twist rule. Part 2 describes experimental testing of this new formula on three plastic-tipped bullets. The new formula is relatively accurate for plastic-tipped bullets whose metal portion has nearly uniform density, but underestimates the gyroscopic stability of bullets whose core is significantly less dense than the jacket.

## Experimental Testing

We ran the numbers using the improved gyroscopic stability formula (Equation 2 of Courtney and Miller 2012) on several plastic-tipped bullet designs to find three that would be well-stabilized at full-power loads and have gyroscopic stability factors approaching 1.0 as the velocity is reduced, taking care to keep the experimental design focused on velocities that remain supersonic to 110 yards where our far chronograph would be for BC measurements and where the target would be to observe key holes.

Figure 3 shows a graph of the predicted gyroscopic stability vs. muzzle velocity for the 60 grain Hornady VMAX at 59 °F and 29.92" Hg (standard conditions) in a 1 in 12" twist barrel for muzzle velocities from 3200 fps down to 1200 fps. The original Miller twist formula (Equation 1) predicts instability over the whole range of muzzle velocities, which is clearly unrealistic since this bullet is known to work well in barrels with 1 in 12" twists. To apply the improved twist formula (Equation 2), we measured the total length and metal length of 10 samples and used average values of the lengths $L = 0.868"$ (total length) and $L_m = 0.738"$ (metal length). The graph shows the gyroscopic stability predicted by the improved formula is just below $S_g$=1.3 for a full-power load and gradually decreases with velocity. Consequently, we expect to see a decrease in BC as velocity is decreased from 3200 fps to 1600 fps, and tumbling to be observed for velocities below 1600 fps.

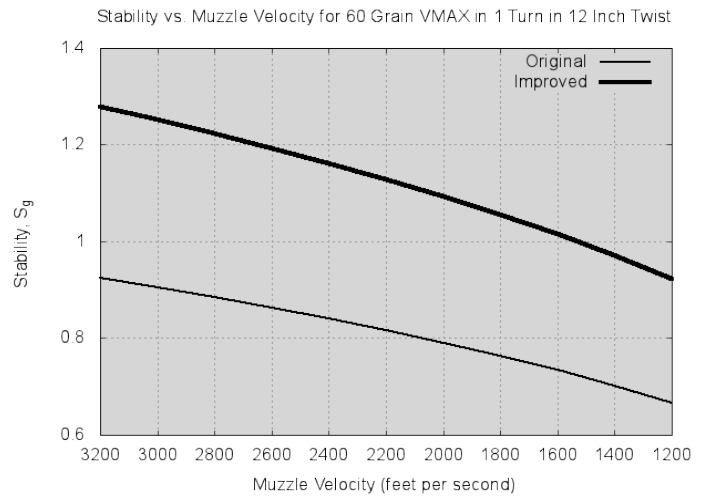

*Figure 3: Gyroscopic stability factor vs. muzzle velocity for the 60 grain VMAX at standard conditions, as predicted by both the original (Equation 1) and improved (Equation 2) twist formulas.*

# A Stability Formula for Plastic-Tipped Bullets

Figure 4 shows a (similar) graph of the predicted gyroscopic stability factor vs. muzzle velocity for the 53 grain Hornady VMAX at standard conditions in a 1 in 12" twist barrel for the same range of muzzle velocities. One ordinarily expects lighter bullets to be significantly more stable than heavier bullets, because they tend to be shorter. However, the 53 grain VMAX is disproportionately long for its weight since Hornady sought to increase the BC with a longer ogive and boat tail. Just as for the 60 grain VMAX, the original Miller twist formula (Equation 1) predicts instability over the whole range of muzzle velocities, which is clearly unrealistic since this bullet is known to work in barrels with 1 in 12" twist rates. Furthermore, Hornady recommends a 1 in 12" twist or faster. To apply the improved twist formula (Equation 2), we used average values of $L = 0.829"$ inches and $L_m = 0.688"$. The graph shows that the gyroscopic stability (according to the improved formula) begins at about 1.36 for a full-power load and gradually decreases with velocity. Consequently, we expect to see stable flight above 1400 fps and tumbling to be observed for velocities below 1400 fps, depending on ambient conditions.

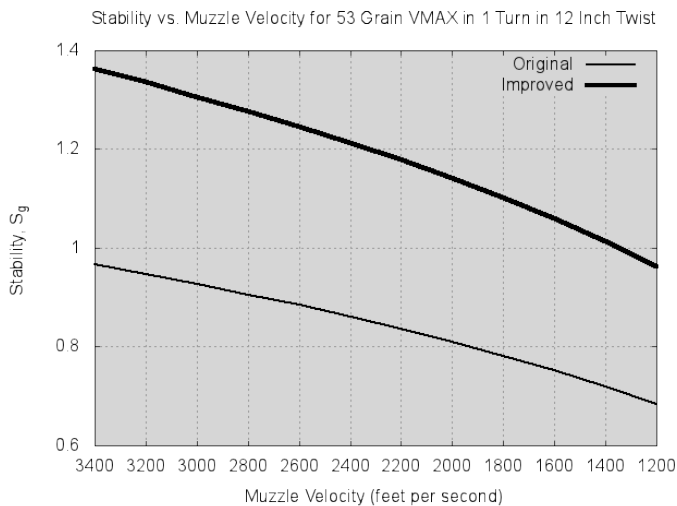

Figure 4: Gyroscopic stability factor vs. muzzle velocity for the 53 grain VMAX at standard conditions, as predicted by both the original (Equation 1) and improved (Equation 2) twist formulas.

Figure 5 shows a graph of the predicted gyroscopic stability factor vs. muzzle velocity for the 40 grain lead-free Nosler Ballistic Tip at standard conditions in a 1 in 12" twist barrel. The composite of the lead-free plastic-tipped bullet is much lighter than lead, and the longer length of the bullet creates less gyroscopic stability than the same-length but heavier lead-core bullets at the same twist rate and conditions. The original Miller twist formula (Equation 1) predicts instability over the whole range of muzzle velocities, which disagrees with Nosler's claim that the bullet works well in barrels with a 1 in 12" twist or faster. To apply the improved twist formula (Equation 2), we used average values of $L = 0.783"$ and $L_m = 0.674"$. The graph shows that the gyroscopic stability factor (according to the improved formula) is below 1.2 for a full-power load and gradually decreases with velocity. Consequently, Equation 2 predicts stable flight for muzzle velocities above 2350 fps and tumbling to be observed for velocities below 2350 fps under standard conditions.

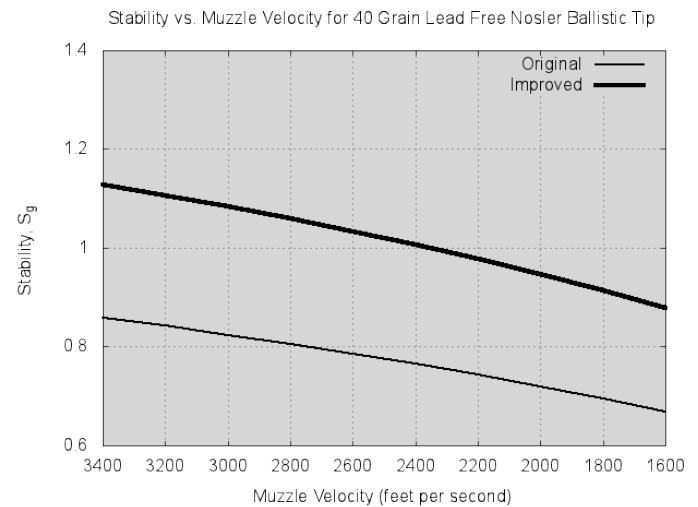

Figure 5: Gyroscopic stability factor vs. muzzle velocity for the 40 grain lead-free Nosler Ballistic Tip at standard conditions, as predicted by both the original (Equation 1) and improved (Equation 2) twist formulas.

The lower density of the powdered-copper core in the 40 grain VMAX may make the assumption of constant density too inaccurate for the formula to yield accurate results. Because the gilding metal jacket is more dense than the copper core, the bullet's axial moment of inertia is much higher than predicted for a bullet of constant (or near constant) density. Consequently, even the improved formula might underestimate gyroscopic stability in this case.

We had hoped to obtain a plastic-tipped copper bullet for testing, but none of our suppliers could provide a tipped .224 TSX in time for testing.

# A Stability Formula for Plastic-Tipped Bullets

The experimental plan was to use two chronographs, one at 30 feet from the muzzle and the other at 330 feet to determine the velocity loss over 300 feet. From this, the JBM BC calculator (www.jbmballistics.com) yields the G1 ballistic coefficient (G1 BC) for a range of muzzle velocities for the three bullets. We would then plot the BC measurements vs. the predicted gyroscopic stability factors. We would also inspect bullet holes carefully and note the percentage of shots that produced key holes at each velocity.

To carry out this plan, MC traveled to a sea level location to conduct the testing, since the ranges near his home are all above 6000 ft. (DM, though near sea level, has no equipment or facilities.) At the reduced atmospheric pressure of high elevations, bullets are much more stable, and it would not be possible to dial down the stabilities from $S_g$=1.3 to below 1.0 and still keep the velocities above the speed of sound. Louisiana Shooters Unlimited (www.louisianashooters.com) kindly offered the use of their range near Lake Charles, Louisiana.

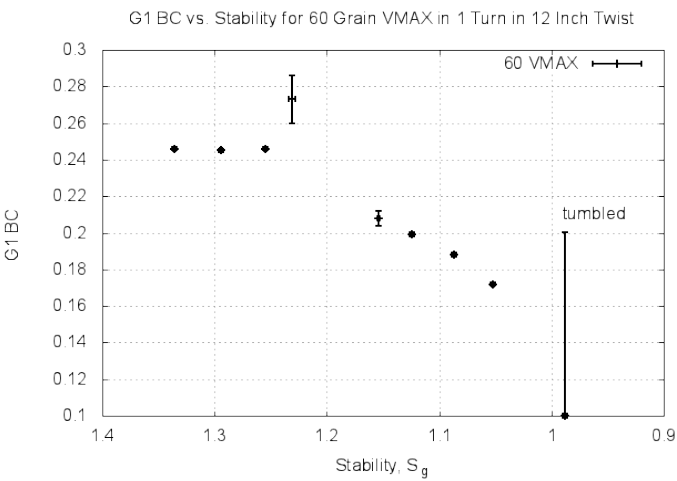

*Figure 6: The G1 BC of the 60 grain VMAX is relatively constant for $S_g$ from 1.34 down to 1.26 before taking a slight jump upward at $S_g$ = 1.21. As $S_g$ is lowered below 1.2, the BC steadily declines until the bullet tumbles for $S_g$ = 0.99, and BC can no longer be meaningfully measured.*

**Results**

Figure 6 shows a graph of the G1 BC vs. $S_g$ for the 60 grain VMAX. Note that for $S_g$ above 1.25, the BC is constant at very close to 0.246. At $S_g$ =1.23, the BC actually increases to close to 0.270, but it demonstrates much larger shot-to-shot variations than at other stabilities. This may be an artifact, because the bullet is "going to sleep" very quickly and flying point forward, but it does exhibit more variation in how fast it goes to sleep than at other stabilities.(Litz 2009b) At gyroscopic stability factors below 1.2, the BC decreased as the stability factor decreased toward 1.0, and at 0.99, the bullets were observed to tumble on 4 of 5 shots (keyholes in the target paper). Therefore, velocity readings were not possible on the far chronograph. The 5[th] shot actually hit (and damaged) the far chronograph and rendered it inoperable, so that it could not be determined whether that 5[th] shot tumbled from instability or only tumbled after it hit the chronograph.

In any event, the experiment demonstrates the accuracy of the improved twist rule because the BC was reduced as expected as the gyroscopic stability was lowered from 1.2 to 1.0, and the bullets were observed to tumble as the stability was lowered below 1.0.

The intent of the experiment was to collect the data to create graphs analogous to Figure 6 for the 53 grain VMAX and 40 grain lead-free Ballistic Tip. However, our BC measurement technique requires two chronographs. Unfortunately, the far chronograph had just been destroyed, and the spare chronograph was 1200 miles away under MC's reloading bench. All we could do with the other two bullets was measure their near velocity, compute their gyroscopic stability according to the improved stability formula, and observe whether they tumbled in flight towards the target at 110 yards.

The results for testing the 53 grain Hornady VMAX are shown in Table 1. The observation of the lack of keyholes is consistent with the predictions of Equation 2, but not Equation 1 for muzzle velocities of 2698 fps down to 1546 fps. However, instability is clearly apparent at 1303 fps (corresponding to $S_g$ = 1.03) because 5 out of 5 bullets were observed to keyhole at the target. This indicates that although Equation 1 is relatively accurate, it should not be considered an absolute predictor of when tumbling of plastic-tipped bullets will be observed.

# A Stability Formula for Plastic-Tipped Bullets

| V (ft/s) | Eqn 1 $S_g$ | Eqn 2 $S_g$ | Keyholes |
|---|---|---|---|
| 1303 | 0.734 | 1.034 | 100% |
| 1546 | 0.777 | 1.094 | 0% |
| 1757 | 0.811 | 1.142 | 0% |
| 1942 | 0.838 | 1.181 | 0% |
| 2123 | 0.863 | 1.216 | 0% |
| 2274 | 0.883 | 1.244 | 0% |
| 2406 | 0.9 | 1.268 | 0% |
| 2564 | 0.919 | 1.295 | 0% |
| 2698 | 0.935 | 1.317 | 0% |
| 2831 | 0.95 | 1.339 | 0% |

Table 1: Stabilities predicted for 53 grain VMAX by Equation 1 and Equation 2 for experimental conditions, along with the percentage of keyholes observed at 110 yards.

| V (ft/s) | Eqn 1 $S_g$ | Eqn 2 $S_g$ | Keyholes |
|---|---|---|---|
| 1717 | 0.714 | 0.939 | 0% |
| 1915 | 0.74 | 0.974 | 0% |
| 2140 | 0.768 | 1.01 | 0% |
| 2319 | 0.789 | 1.038 | 0% |
| 2502 | 0.81 | 1.064 | 0% |
| 2665 | 0.827 | 1.087 | 0% |
| 2806 | 0.841 | 1.106 | 0% |
| 2999 | 0.86 | 1.131 | 0% |
| 3157 | 0.875 | 1.15 | 0% |

Table 2: Stabilities predicted for 40 grain lead-free Nosler Ballistic Tip predicted by Equation 1 and Equation 2 for experimental conditions, along with the percentage of keyholes observed at 110 yards.

The results for testing the 40 grain lead free Nosler Ballistic Tip are shown in Table 2. Note that the original Miller stability formula predicts instability over the whole velocity range; however, no keyholes were observed at any velocity tested.

Also note that the improved formula (Equation 2) also fails to predict the stability of the bullet at 1915 fps and 1717 fps, predicting instability when the bullet is, in fact, stable. This can be explained by the composite construction of the bullet. The bullet jacket is gilding metal, which is much heavier than the powdered-copper core. Consequently, the low core density produces a bullet that is relatively light for its length. More importantly, the higher density of the jacket (relative to the core) increases the ratio of the axial moment of inertia to the transverse (tumbling) moment of inertia. Therefore, the gyroscopic stability is greater than that predicted by Equation 1, which is based on the assumption that the metal portion of the bullet has relatively uniform density throughout its volume.

## Discussion

Differing bullet constructions make it challenging to predict bullet gyroscopic stability with simple empirical formulas. However, just as the original Miller twist rule was a significant improvement over the Greenhill formula for bullets of relatively uniform density, the new formula (Equation 2) for plastic-tipped bullets with relatively uniform density in their metal portion is a significant improvement over the original Miller stability formula. However, the gyroscopic stability of bullets with a powdered core that is significantly less dense than the jacket, is underestimated by Equation 2, and probably results from the effect of the hard-to-measure moments of inertia. Further planned experiments with such bullets may suggest improvements in either Equation 1 or Equation 2.

## Bibliography

**Courtney, Michael and Miller, Don.** A Stability Formula for Plastic-Tipped Bullets: Part 1. *Precision Shooting.* Jan 2012.

**Litz, Brian.** *Applied Ballistics for Long Range Shooting.* Cedar Springs, MI : Applied Ballistics, LLC, 2009a, 2nd Edition, 2011.

**Litz, Brian.** *Accurate Specifications of Ballistic Coefficients.* The Varmint Hunter Magazine, Issue 72. 2009b.

**McDonald, William and Algren, Ted.** *Sierra Loading Manual, 5th ed.*, Section 2.5. Examples of Ballistic Coefficient Measurements.

**McDonald, William and Algren, Ted.** *Sierra Loading Manual, 4th ed.*, Section 4.6. Ballistic Coefficient Dependence on Coning Motion.

**Miller, Don.** A New Rule for Estimating Rifling Twist: An Aid to Choosing Bullets and Rifles. *Precision Shooting. March 2005, pp. 43-48.*

**Miller, Don.** How Good Are Simple Rules for Estimating Rifle Twist. *Precision Shooting. June 2009, pp. 48-52.*

***Note:*** The addresses and affiliations listed are where the work was completed in 2010-2012.   Michael Courtney's current affiliation is BTG Research, Baton Rouge, Louisiana.